\documentclass[11pt,a4paper]{emulateapj}
 \usepackage{url}
\usepackage{amsmath}
\usepackage{xcolor} 
\usepackage{graphicx}
\usepackage[export]{adjustbox}
\usepackage{psfig}
\usepackage{multirow}
\usepackage[flushleft]{threeparttable}
\usepackage[symbol*]{footmisc}
\DefineFNsymbolsTM{myfnsymbols}{
 \textasteriskcentered *
  \textdagger    \dagger
  \textsection   \mathsection
  \textbardbl    \|%
  \textparagraph \mathparagraph 
}%
\setfnsymbol{myfnsymbols}
 \begin{document}

\title{Buoyant AGN bubbles in the quasi-isothermal potential of NGC~1399}

\author{Yuanyuan Su\altaffilmark{$\ddagger$1}}
\author{Paul E.\ J.\ Nulsen\altaffilmark{1}}
\author{Ralph P. Kraft\altaffilmark{1}}
\author{William R. Forman\altaffilmark{1}}
\author{Christine Jones\altaffilmark{1}}
\author{Jimmy A. Irwin\altaffilmark{2}}
\author{Scott W. Randall\altaffilmark{1}}
\author{Eugene Churazov\altaffilmark{3}}
\affil{$^1$Harvard-Smithsonian Center for Astrophysics, 60 Garden Street, Cambridge, MA 02138, USA}
\affil{$^2$Department of Physics and Astronomy, University of Alabama, Box 870324, Tuscaloosa, AL 35487, USA}
\affil{$^3$Max Planck Institute for Astrophysics, Karl-Schwarzschild-Str. 1, 85741, Garching, Germany}

\altaffiltext{$\ddagger$}{Email: yuanyuan.su@cfa.harvard.edu}

\keywords{
X-rays: galaxies: luminosity --
galaxies: ISM --
galaxies: elliptical and lenticular  
Clusters of galaxies: intracluster medium  
}

\begin{abstract}

The Fornax Cluster is a low-mass cool-core galaxy cluster. We present a deep {\sl Chandra} study of NGC~1399, the central dominant elliptical galaxy of Fornax. The cluster center harbors two symmetric X-ray cavities coincident with a pair of radio lobes fed by two collimated jets along a north-south axis. A temperature map reveals that the AGN outburst has created a channel filled with cooler gas out to a radius of 10 kpc. The cavities are surrounded by cool bright rims and filaments that may have been lifted from smaller radii by the buoyant bubbles. X-ray imaging suggests a potential ghost bubble of $\gtrsim$ 5\,kpc diameter to the northwest. We find that the amount of gas lifted by AGN bubbles is comparable to that which would otherwise cool, demonstrating that AGN driven outflow is effective in offsetting cooling in low-mass clusters. The cluster cooling time scale is $>30$ times longer than the dynamical time scale, which is consistent with the lack of cold molecular gas at the cluster center. The X-ray hydrostatic mass is consistent within 10\% with the total mass derived from the optical data. The observed entropy profile rises linearly, following a steeper slope than that observed at the centers of massive clusters; gas shed by stars in NGC~1399 may be incorporated in the hot phase. However, it is far-fetched for supernova-driven outflow to produce and maintain the thermal distribution in NGC~1399 and it is in tension with the metal content in the hot gas.

\end{abstract}

\section{\bf Introduction}

Active galactic nuclei (AGN) 
are the engines of the cooling and heating cycles taking place at the centers of galaxy clusters. The accretion of cluster gas onto the supermassive black hole of the brightest cluster galaxy (BCG) triggers outflow and releases energy to quench cooling flows. 
The observational evidence of AGN feedback is primarily manifested by the presence of X-ray cavities coincident with radio lobes, which were previously detected in the BCGs of Perseus and Virgo using {\sl ROSAT} (Churazov et al.\ 2000; Churazov et al.\ 2001). 
Through hydrodynamic simulations, Churazov et al.\ (2001) illustrate that bubbles inflated by radio jets from an earlier AGN outburst rise through the intracluster medium (ICM) and displace the ambient X-ray emitting gas.
The {\sl Chandra} X-ray observatory has revealed that such phenomena are pervasive in the cool-core clusters (e.g., McNamara \& Nulsen 2007; Nulsen et al.\ 2013; McNamara et al.\ 2000; Kraft et al.\ 2000; Finoguenov \& Jones et al.\ 2001).  
The power deposited via radio jets into
radio lobes is more than enough to prevent the ICM from condensing into cold gas and forming stars, which it would do
in the absence of heating.
A crucial link in this feedback loop, the exact process of the energy dissipation from the jet power into the thermal emission, remains to be identified. To this end, shock waves, turbulent ICM, and uplifted low-entropy gas have been observed at the centers of cool core clusters and these mechanisms have been proposed as a means of expending the jet power (Randall et al.\ 2015; Zhuravleva et al.\ 2015; Kirkpatrick \& McNamara 2015). All
these processes are affected by the poorly understood microphysics of
the cluster plasma physics on a microscopic scale (e.g., Nulsen et al.\ 2002; Su et al.\ 2017).

{\sl Chandra} has greatly enhanced our understanding of the interplay of AGN feedback and hot gas, thanks to observations of many galaxy clusters with its superb spatial resolution. 
Nevertheless, studies of AGN outbursts have been preferentially performed for massive clusters and/or radio bright BCGs.
In this paper, we present a case study of the nearest cluster in the southern sky, the Fornax Cluster. 
It is a low-mass cool-core cluster with an average ICM temperature of $T_{X}=1.3$\,keV as revealed by previous {\sl ROSAT} observations (Jones et al.\ 1997; Rangarajan et al.\ 1995). We derive a total mass of $5\times10^{13}$\,M$_{\odot}$ within $R_{500}$\footnote{$R_{500}\approx 0.6 r_{\rm vir}$ is the radius within which the average density is 500 times the critical density of the Universe.}$\approx315$\,kpc from the $M_{500}-T_{X}$ relation calibrated for galaxy groups (Lovisari et al.\ 2015).
Using the Very Large Array (VLA), Killeen et al.\ (1988) found that
the radio source in the BCG, NGC~1399, consists of narrow opposed jets feeding
extended lobes with a very low radio luminosity ($\sim10^{39}$\,erg\,s$^{-1}$).  
Paolillo et al.\ (2002), Shurkin et al.\ (2008), and Dunn et al.\ (2010) detected a pair of X-ray
cavities corresponding to these two radio lobes. 
We revisit the effects of AGN outbursts in the ICM using deep {\sl Chandra} observations.
Thanks to the proximity ($<20$\,Mpc) of NGC~1399, its gas properties can be studied in great detail, revealing features that would otherwise be missed.

We adopt the redshift of $z=0.00475$ for NGC~1399 from the NASA/IPAC Extragalactic Database
and a luminosity distance of 19 Mpc ($1^{\prime} = 5.49$ kpc) taken from Paolillo et al.\ (2002). 
The observations and data reduction are described in \S2. Further analysis and the results are presented in \S3.
The implications of gas cooling and heating are discussed in \S4, and our main conclusions are summarized in \S5. Uncertainties reported are quoted at a confidence level of 68\% throughout this work.

\section{\bf observations and data reductions}

A total of 250 ksec {\sl Chandra} observations centered on NGC~1399 are included in this study; the observational log is listed in Table~1. {\sl CIAO}~4.8 and {\sl CALDB}~4.6.9 were used for the {\sl Chandra} data reduction.   
All the observations were reprocessed from level 1 events using the {\sl CIAO} tool {\tt chandra\_repro} to guarantee the latest, consistent calibrations. 
We screened background flares beyond 3$\sigma$ using the light curve filtering script {\tt lc\_clean}; net exposure times are listed in Table~1. Readout artifacts were subtracted from both imaging and spectral analysis. Point sources were detected in a 0.3--7.0 keV image with {\tt wavdetect}, supplied with a 1.0 keV PSF map. The detection threshold was set to 10$^{-6}$. The scales ranged from 1 to 8 pixels, increasing in steps of a factor of $\sqrt{2}$.

\subsection{Imaging analysis}

\begin{figure*}
   \centering
    \includegraphics[width=0.45\textwidth]{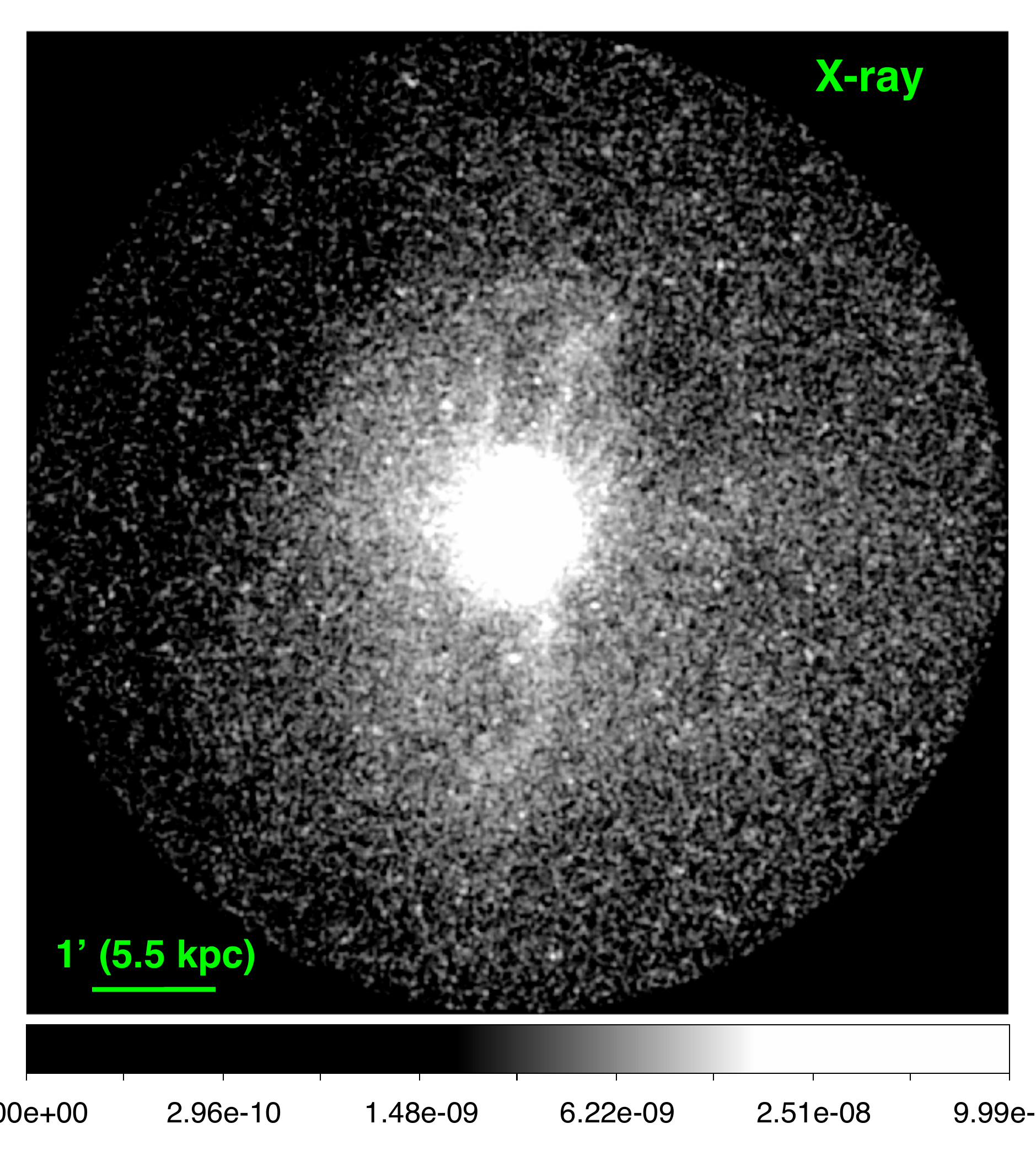}
            \includegraphics[width=0.45\textwidth]{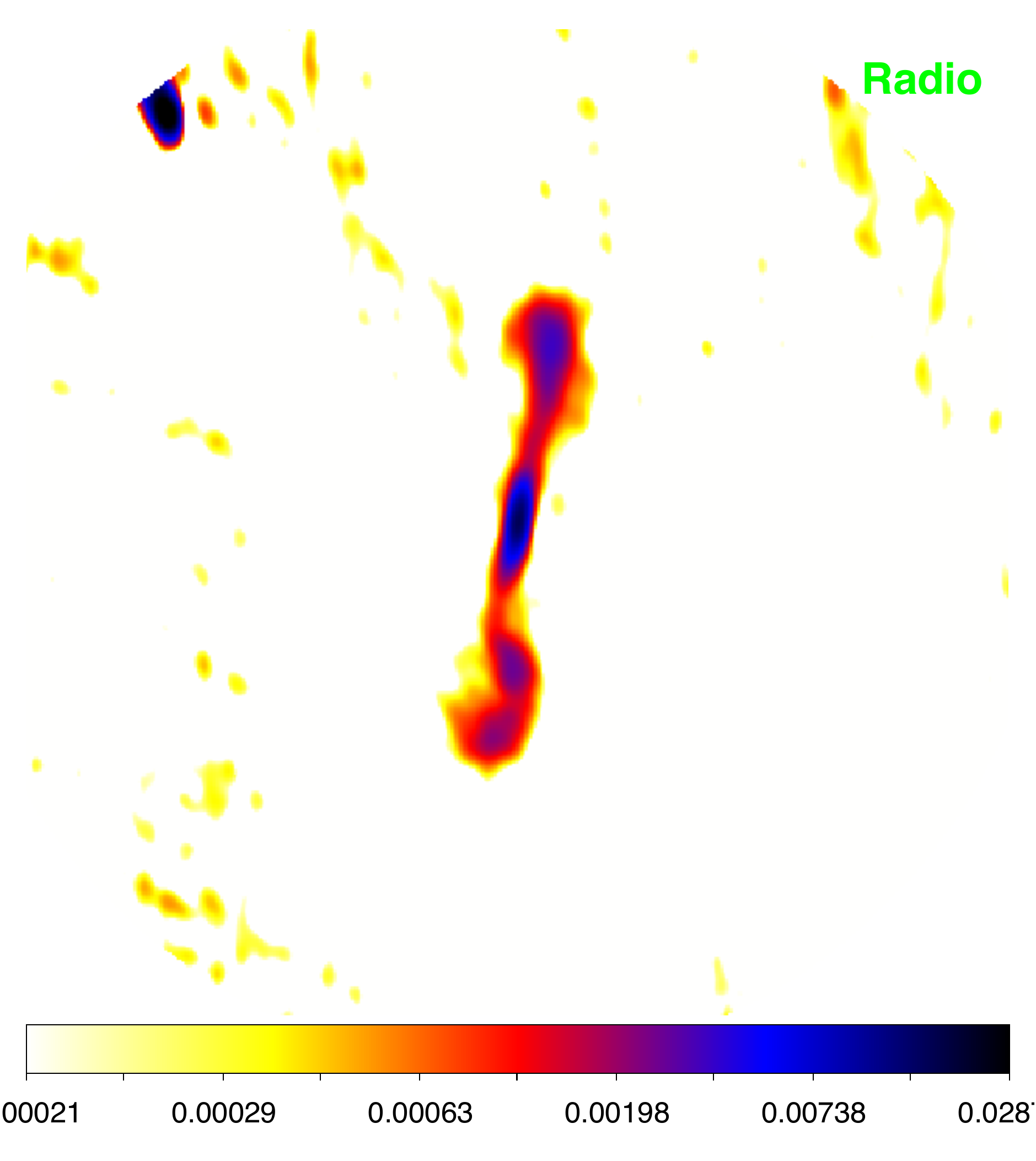}
      \includegraphics[width=0.45\textwidth]{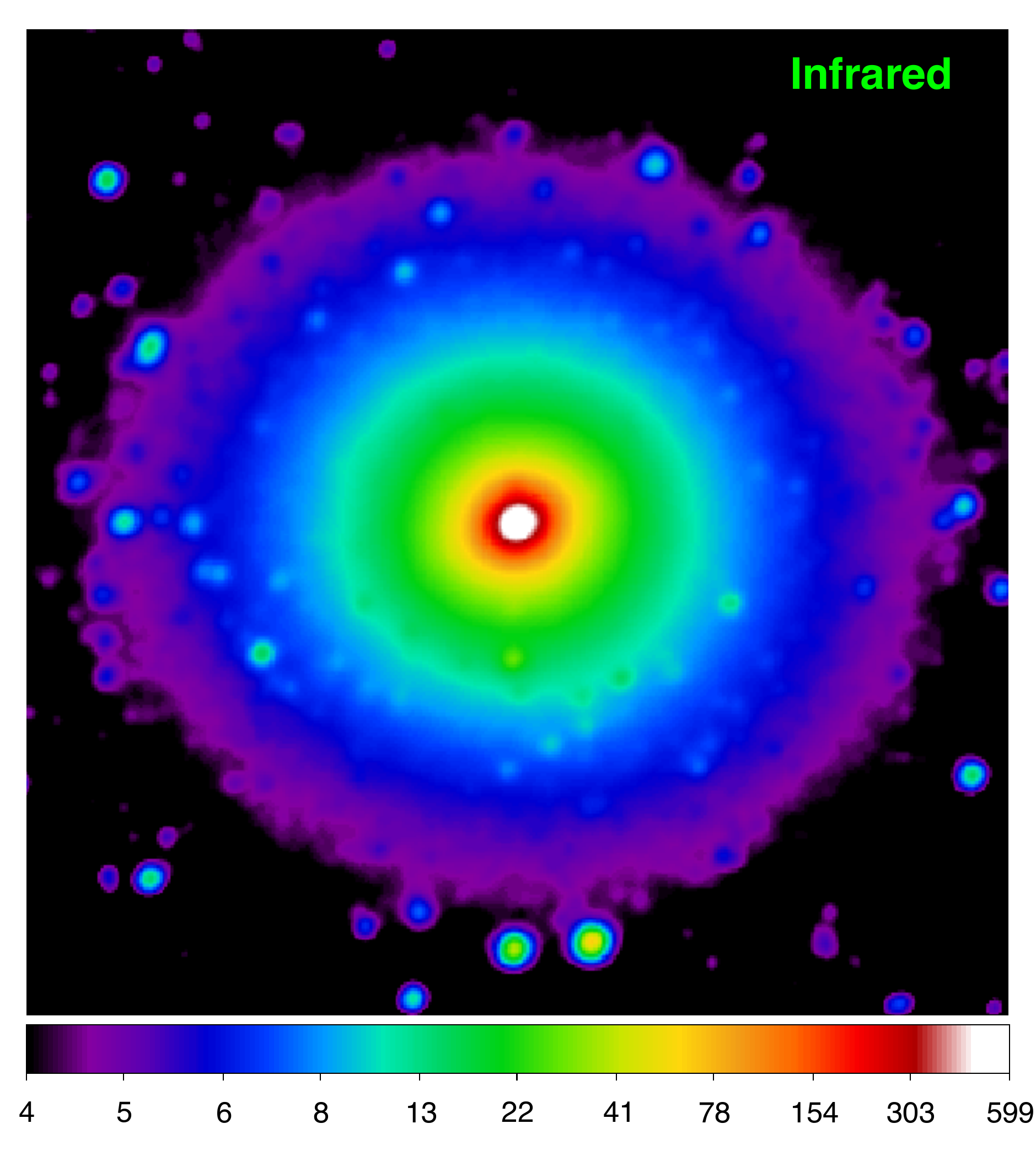}
     \includegraphics[width=0.45\textwidth]{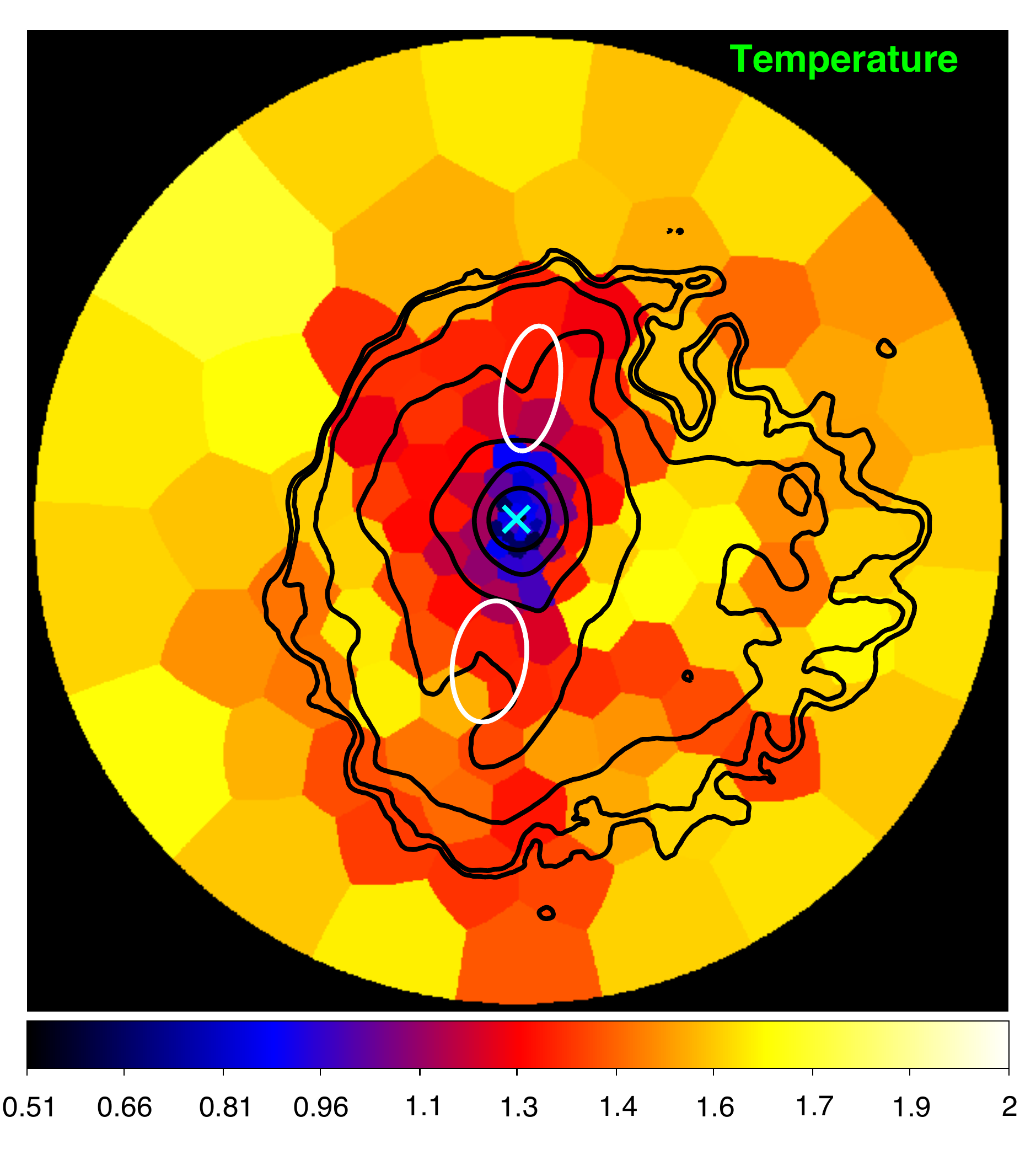}
\figcaption{\label{fig:img} {\it top-left}: {\sl Chandra} X-ray image of NGC~1399 in the 0.5--2.0 keV energy band in units of photon\,cm$^{-2}$\,s$^{-1}$ per pixel ($0\farcs492\times0\farcs492$). The image was exposure corrected and background subtracted. 
{\it top-right}: {VLA} 4.88 GHz radio image of NGC~1399 in units of Jy beam$^{-1}$. 
{\it bottom-left}: WISE 3.4 $\mu$m\, image of NGC~1399. {\it bottom-right}: 
Two-dimensional temperature distribution of the hot gas in NGC~1399 in units of keV. Cyan cross: X-ray peak. White ellipses: bubble locations. Black contours: {\sl Chandra} X-ray emission in the 0.5--2.0 keV energy band. All these images are on the same physical scale.}
\end{figure*}

We extracted images from each event file in 7 energy bands: 0.5--0.7 keV, 0.7--0.9 keV, 0.9--1.1 keV, 1.1--1.3 keV, 1.3--1.5 keV, 1.5--1.7 keV, and 1.7--2.0 keV. Each image was normalized with a monochromatic exposure map defined at the central energy of each band. We subtracted an approximation of the background from each image using the blank-sky fields available in the {\sl CALDB}. The background level was scaled by the count rate in the 9.5--12.0 keV energy band relative to the observation. We replaced resolved point sources with pixel values interpolated from surrounding background regions using the {\sl CIAO} tool {\tt dmfilth}. A final 0.5--2.0 keV image of NGC~1399 was produced by adding all 7 narrow band images and presented in Figure~\ref{fig:img}-top-left. All but one observation were taken with ACIS-S. Therefore, we restrict the field-of-view to the inner 9$^{\prime}$ centered on the X-ray peak (03h38m29s, -35d27m01.1s).

 \begin{deluxetable}{cccccc}
\tablewidth{0pc}
 \centering
\tablecaption{Chandra observational log for the analyses of NGC~1399}
\tablehead{
\colhead{Obs ID}&\colhead{Date}&\colhead{Instrument}&\colhead{exp (ks)}&\colhead{RA (deg)}&\colhead{Dec (deg)}}
\startdata
319&2000-01&ACIS-S&56&54.62&-35.45\\
4172&2003-05&ACIS-I&45&54.61&-35.43\\
9530&2008-06&ACIS-S&60&54.62&-35.45\\
14527&2013-07&ACIS-S&28&54.62&-35.45\\
14529&2015-11&ACIS-S&32&54.62	&-35.45\\
16639&2014-10&ACIS-S&30&54.62	&-35.45
\enddata
\end{deluxetable}

\subsection{Spectral analysis}

We extracted spectra for regions of interest from the ACIS-S3 and ACIS-I CCDs. 
Response files and matrices were produced for each spectrum using the CIAO tools {\tt mkwarf} and {\tt mkacisrmf}, respectively.  All spectra were grouped to have at least one count per energy bin. Spectral fitting was performed with {\sl XSPEC} 12.7 using the C-statistic. The energy range for spectral fitting was restricted to 0.5--7.0 keV.  
Photoionization cross-sections were taken from Balucinska-Church \& McCammon (1992). 
We adopted a Galactic 
hydrogen column of $N_{\rm H}=1.5\times10^{20}$ cm$^{-2}$ toward NGC~1399, which was deduced 
from the LAB map (Kalberla et al.\ 2005) incorporated in the {\sl HEASARC} $N_{\rm H}$ tool. We use {\tt phabs} to model the foreground absorption. Blank-sky background normalized to the hard band of each observation was applied unless stated otherwise. 
We use the thermal emission model ${\tt vapec}$ to model the hot gas component; the abundances of He, C, and N were fixed at the solar values; the abundances of O, Ne, Mg, Si, S, Fe, and Ni were allowed to vary freely unless stated otherwise; all other elements were linked to Fe. The solar abundance standard of Asplund at al.\ (2006) was adopted. 
Unresolved low mass X-ray binaries (LMXB) also contribute to the diffuse X-ray emission. 
We use a power law model with an index of 1.6, ${\tt pow}_{1.6}$, to describe the unresolved LMXB (Irwin et al.\ 2003) and we allow its normalization to vary.

\section{\bf analysis and results}

\begin{figure*}
   \centering
           \includegraphics[width=0.45\textwidth]{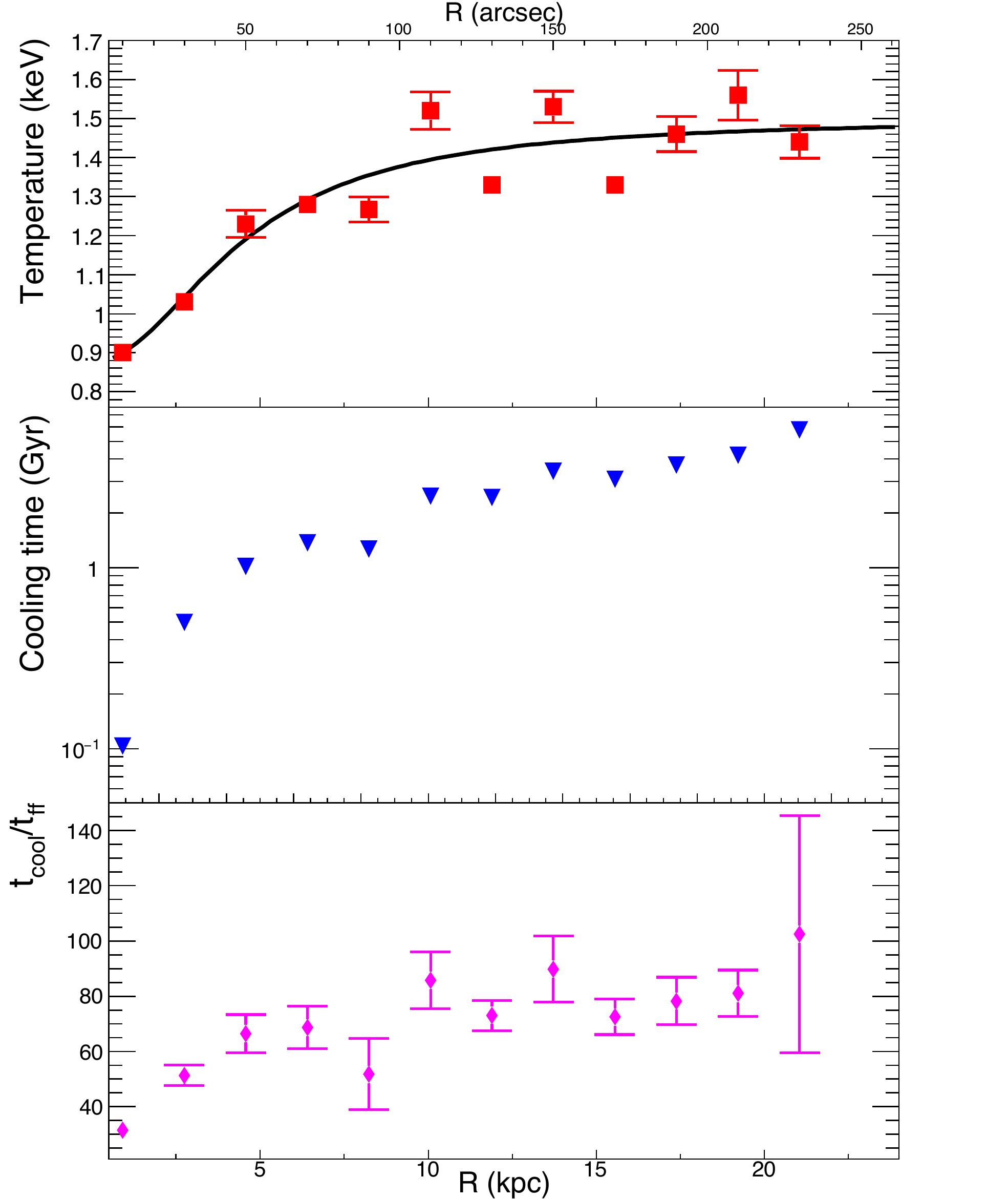}
    \includegraphics[width=0.45\textwidth]{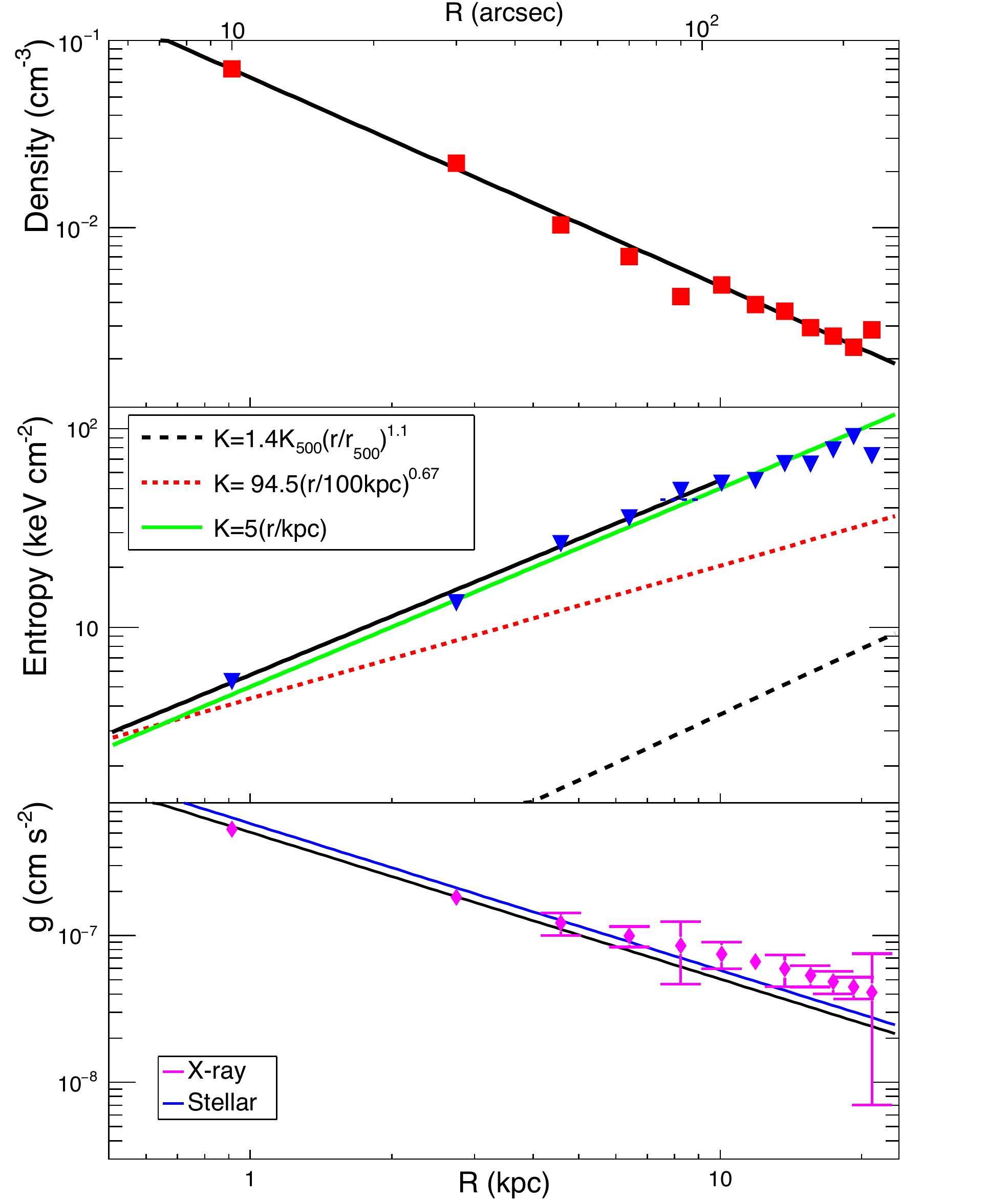}

\figcaption{\label{fig:pro}Deprojected radial profiles of temperature (left-top), electron density (right-top), cooling time (left-middle), entropy (right-middle), $t_{\rm cool}/t_{\rm ff}$ (left-bottom), and the gravitational acceleration ($g$) (right-bottom) centered  on NGC~1399. The best-fit temperature, density, entropy, and $g$ profiles are plotted in black solid lines. 
{\it right-middle}: We compare the entropy profile to the power-law profile predicted by ``adiabatic" cosmological simulations that account only for gravity (black dashed line, Voit et al. 2005), the average entropy profile measured at the center of massive clusters (red dashed line, Panagoulia et al.\ 2014), and the entropy profile for single-phase elliptical galaxies (green line, Voit et al.\ 2015). {\it right-bottom}: We compare the gravitational acceleration to that expected from stellar velocity dispersion (blue).
Statistical error bars are smaller than the symbol size for some data sets.}
\end{figure*}

\subsection{Deprojected radial profiles of gas properties}

We extract spectra from a set of 12 concentric annuli ranging from the central X-ray peak to a radius of $4^{\prime}$ (22\,kpc). The radial width of each annulus is $20^{\prime\prime}$. 
We use the {\sl XSPEC} mixing model {\tt projct} to perform the deprojection analysis. 
{We fit the spectra with the model: ${\tt projct}\times({\tt phabs}\times{\tt vapec})+{\tt phabs}\times{\tt pow}_{1.6}$. We obtain 
a C-statistics of 6809.91 for 5912 degrees of freedom. The results are tabulated in Table~2.}
The best-fit deprojected temperature profile is shown in Figure~\ref{fig:pro}-top-left.
The temperature is below 1\,keV at the cluster center and rises to 1.5\,keV beyond 10\,kpc. 
We fit the deprojected temperature profile to a simplified version of the universal profile of cluster cool cores described by Allen et al.\ (2001). 
\small
\begin{equation}\label{eq:1}
T_{\rm 3D}(r)=T_0\frac{(r/r_c)^2+T_m/T_0}{{(r/r_c)}^2+1}
\end{equation}
\normalsize
and we find $T_0=1.50\pm0.05$\,keV, $r_c=4.53\pm1.33$\,kpc, $T_m=0.87\pm0.09$\,keV.
We derive the density of each shell through the best-fit value of the normalization, using
\begin{equation}\label{eq:2}
norm=\frac{10^{-14}}{4\pi {D_A}^2(1+z)^2}{n_e n_H}{V},
\end{equation}
where $D_A$ is the angular diameter distance of the source, $V$ is the volume of each concentric spherical shell, $n_e$ and $n_H$ are
the electron and proton densities, for which we assume $n_e=1.21n_H$. 
The resulting deprojected density profile is presented in Figure~\ref{fig:pro}-top-right. 
We fit it to a power-law model of
\begin{small}
\begin{equation}\label{eq:3}
n_e(r)={n_{0}}{(r/{\rm 1\,kpc})}^{-\alpha}
\end{equation}
\end{small}
and we obtain 
$n_0=0.063\pm0.001$\,cm$^{-3}$ and $\alpha=1.11\pm0.01$.
These deprojected results were used to derive the three-dimensional entropy $K=kT\,{n_e}^{-2/3}$. As shown in Figures~\ref{fig:pro}-middle-right, the entropy stays below 50\,keV\,cm$^{-2}$ within a radius of 10\,kpc; the profile declines all the way to the cluster center and no central ``entropy floor" was observed. The best-fit slope of the entropy profile within 10\,kpc is $1.01\pm0.02$,
following a steeper slope than that observed at the centers of more massive clusters ($\propto r^{0.67}$ Panagoulia et al.\ 2014). 
Its normalization is 10 times higher than that predicted by ``adiabatic'' cosmological simulations for purely gravitational collapse (Voit et al.\ 2005). 
This discrepancy indicates that the hot gas in NGC~1399 may have a different origin from 
the vast ICM and the BCGs of more massive clusters.
The single temperature Fe abundance varies in the range of 0.5--1.0\,Z$_{\odot}$ and does not show any obvious gradient.

 \begin{deluxetable*}{ccccccc}
\tablewidth{-0pc}
 \centering
\tablecaption{Results of the azimuthally-averaged deprojected analysis of NGC~1399}
\tablehead{
\colhead{Radius}&\colhead{T}&\colhead{$n_e$}&\colhead{Fe}&\colhead{${L_{X}}^{*}$}&\colhead{${M_{X, \rm tot}}^{*}$}\\
\colhead{(kpc)}&\colhead{(keV)}&\colhead{($10^{-3}$\,cm$^{-3}$)}&\colhead{(Z$_{\odot}$)}&\colhead{($10^{40}$\,erg\,s$^{-1}$)}&\colhead{($10^{10}$\,M$_{\odot}$)}}
\startdata
0.915&$0.91^{+0.01}_{-0.01}$& $58.30^{+1.03}_{-1.11}$ &$0.69^{+0.01}_{-0.02}$&7.34&$3.84\pm0.10$\\
2.745&$1.03^{+0.01}_{-0.01}$&$18.22^{+0.68}_{-0.62}$ &$0.55^{+0.03}_{-0.04}$&11.24&$10.00\pm1.24$\\
4.575&$1.23^{+0.03}_{-0.02}$&$8.53^{+0.42}_{-0.40}$ &$0.52^{+0.10}_{-0.10}$&14.09&$18.10\pm3.21$\\
6.405&$1.28^{+0.04}_{-0.02}$&$5.79^{+0.33}_{-0.55}$&$0.49^{+0.14}_{-0.14}$&17.01&$28.74\pm4.66$\\
8.235&$1.27^{+0.03}_{-0.02}$&$4.02^{+0.40}_{-0.30}$ &$0.62^{+0.23}_{-0.21}$&20.16&$40.71\pm18.53$\\
10.065&$1.56^{+0.06}_{-0.04}$&$4.17^{+0.21}_{-0.24}$ &$0.83^{+0.16}_{-0.18}$&23.47&$53.16\pm10.84$\\
11.895&$1.31^{+0.02}_{-0.02}$&$3.01^{+0.16}_{-0.15}$ &\multirow{ 2}{*}{$0.95^{+0.05}_{-0.13}$}&26.68&$65.69\pm7.33$\\
13.725&$1.50^{+0.06}_{-0.06}$&$2.80^{+0.18}_{-0.12}$&&29.94&$78.17\pm18.99$\\
15.555&$1.33^{+0.02}_{-0.03}$&$2.35^{+0.13}_{-0.13}$&\multirow{ 2}{*}{$0.77^{+0.14}_{-0.12}$}&33.25&$90.56\pm15.07$\\
17.385& $1.43^{+0.08}_{-0.10}$&$2.49^{+0.16}_{-0.11}$&&36.67&$102.86\pm17.62$\\
19.215&$1.52^{+0.10}_{-0.10}$&$1.67^{+0.16}_{-0.16}$&\multirow{ 2}{*}{$1.07^{+0.15}_{-0.16}$}&40.08&$115.07\pm19.43$\\
21.045&$1.51^{+0.06}_{-0.06}$ &$2.36^{+0.13}_{-0.13}$&&43.47&$127.21\pm105.38$
\enddata
\tablecomments{*: Enclosed bolometric X-ray luminosity (0.01--100\,keV) and hydrostatic mass.}
\end{deluxetable*}


We calculate the gas cooling time, $t_{\rm cool}$, using
 \begin{equation}\label{eq:4}
 t_{\rm cool}=\frac{3P}{2n_en_H\Lambda(T,Z)},
\end{equation}
where $P=1.8n_ekT$ is the pressure and $\Lambda(T,Z)$ is the cooling function determined by the plasma temperature and metallicity.
The cooling time and gas entropy in NGC~1399 is below 0.5\,Gyr and 10\,keV\,cm$^{2}$ respectively at the center of NGC~1399, (Figure~\ref{fig:pro}), indicating that Fornax is a cool-core cluster. The cooling radius is conventionally defined as the radius within which the cooling time is shorter than 1\,Gyr (strong cool core) or the lookback time to $z=1$ (7.7\,Gyr) (weak cool core) (Hudson et al.\ 2010). For NGC~1399, these correspond to cooling radii of 4.5\,kpc and 25\,kpc, respectively.  
This cool core is quite small, which may be a consequence of its small cluster mass.
  
The best-fit temperature and density profiles allow us to derive a smooth pressure profile, $P$. Assuming hydrostatic equilibrium and spherical symmetry,
the gravitational acceleration, $g$, can be related to $P$ as
\begin{equation}\label{eq:5}
g=\frac{d \Phi}{d r}=-\frac{1}{\rho}\frac{dP}{dr},
\end{equation}
{where $\Phi$ is the gravitational potential and $\rho=n\mu m_p$ with $n=1.91n_e$ the particle density, $\mu=0.61$ the mean particle
weight, and $m_p$ the proton mass.}
If the potential is isothermal, then we expect the circular velocity,
\begin{equation}\label{eq:6}
{v}=\sqrt{\frac{GM}{r}}=\sqrt{gr}
\end{equation}
to be constant. 
As shown in Figure~\ref{fig:pro}-right-bottom, we obtain a best-fit $v$ of $394\pm63$\,km\,s$^{-1}$, which is $\lesssim10\%$ smaller than the expectation from the optical data $v_{*}=\sqrt{2}\sigma_{*}=$421\,km\,s$^{-1}$, where $\sigma_{*}$ is the average one-dimensional stellar velocity dispersion taken from Saglia et al.\ (2000). 
We then determine the free fall time $t_{\rm ff}=\sqrt{2r/g}$. As shown in Figure~\ref{fig:pro}-bottom-left, the cooling time is more than 30 times the free fall time over the entire cluster center.

\subsection{X-ray cavities}

\begin{figure}
   \centering
    \includegraphics[width=0.5\textwidth]{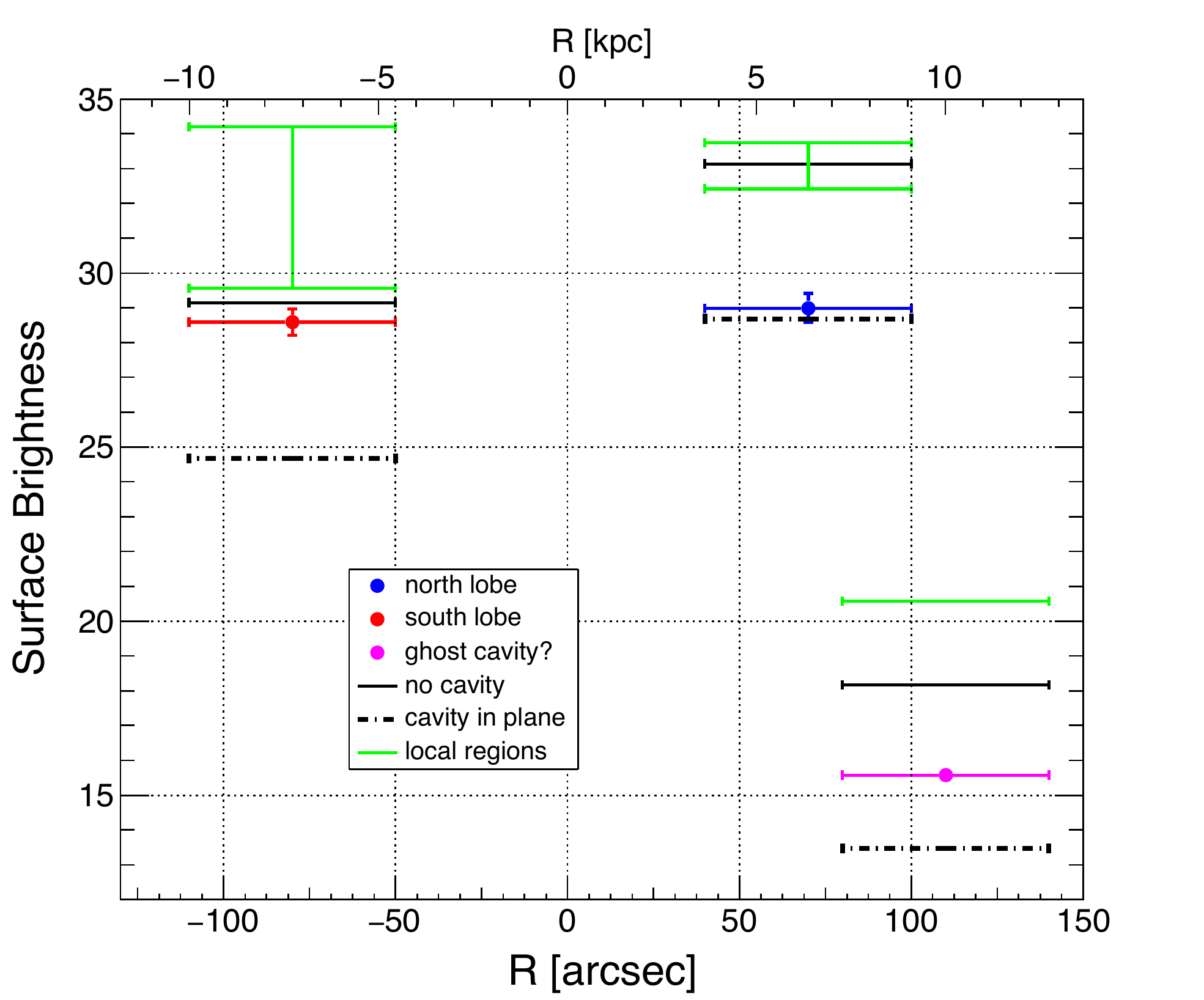}    
\figcaption{\label{fig:cavity} The measured X-ray surface brightness in the 0.5--2.0 keV energy band and in units of $10^{-9}$\,photon\,cm$^{-2}$\,s$^{-1}$\,arcsec$^{-2}$ of the northern lobe (blue), the southern lobe (red), and the ghost cavity candidate (magenta), marked in solid lines. Those of regions at the same radius adjacent to the cavities are indicated in the green lines. Dash-dot lines: the expected values if the cavities are in the plane of the sky. Black solid line: the expected values if there are no cavities. X-axis is their projected distance to the nucleus.}
\end{figure}

\begin{figure}
   \centering
    \includegraphics[width=0.5\textwidth]{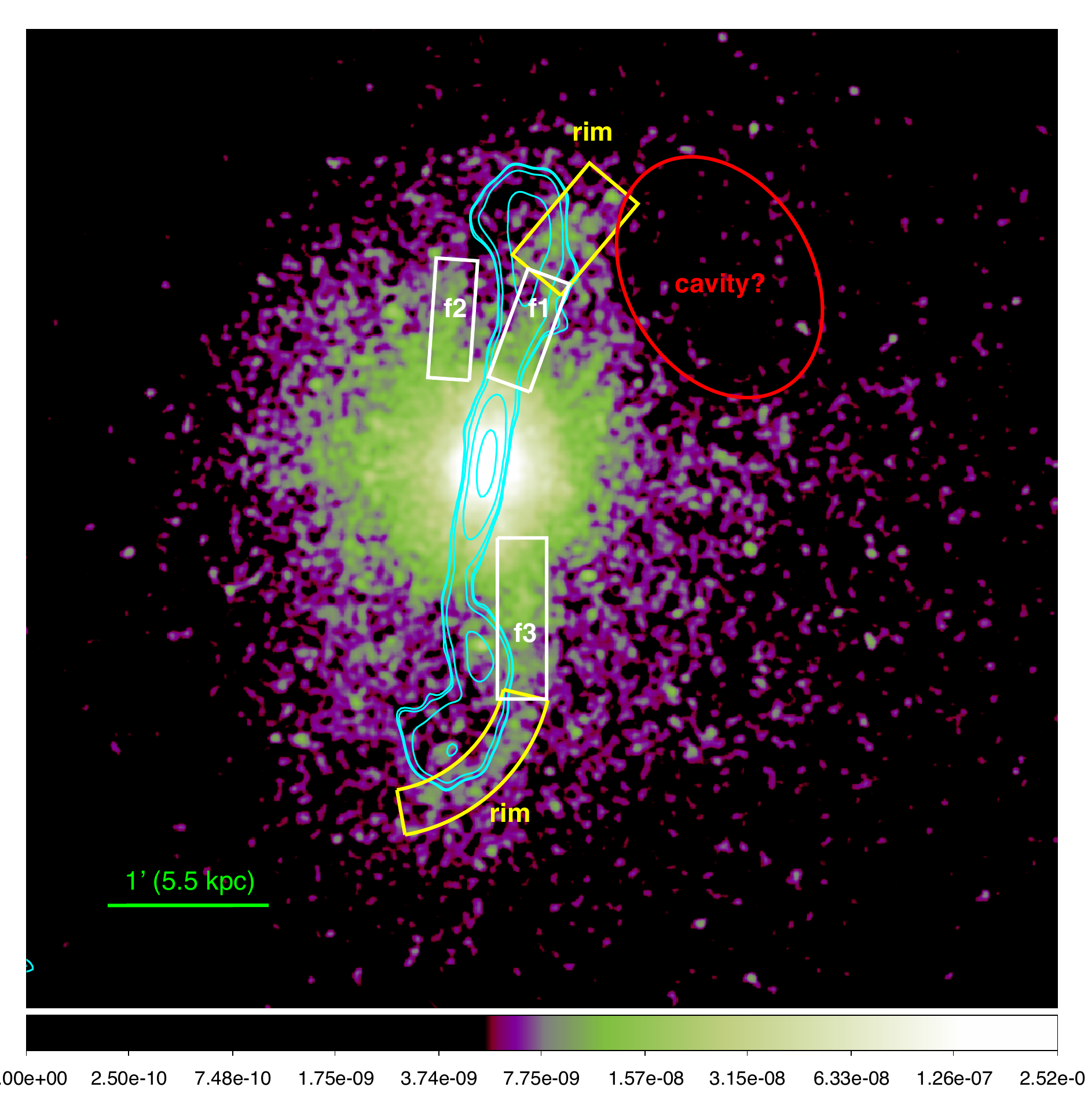}    
\figcaption{\label{fig:N1399} {\sl Chandra} X-ray image of NGC~1399 in the 0.5--2.0 keV energy band in units of photon\,cm$^{-2}$\,s$^{-1}$. Cyan: VLA radio contour levels are set at [5, 5.4, 7.9, 22, 100]$\times$$\sigma_{\rm rms}$ where $\sigma_{\rm rms}$=0.1 mJy beam$^{-1}$. Yellow: uplifted rims; White: uplifted filaments (see \S3.4 and Table~3).}
\end{figure}

Images of NGC~1399 at multiple wavelengths are presented in Figure~\ref{fig:img}. A pair of symmetric cavities centered at radii of 6.7\,kpc to the north and 7.1\,kpc to the south are visible in the X-ray image; their surface brightness decrement relative to the neighboring regions is shown in Figure~\ref{fig:cavity}. They are spatially correlated with the two radio lobes
that are inflated by jets from the nucleus. 
We note that the stellar
distribution of NGC 1399 is elongated in the east-west direction with an ellipticity of 0.1--0.2 (Dirsch et al.\ 2003),
perpendicular to the axis of the outburst, consistent with AGN-driven
outflows occurring preferentially along the minor axis of the host
galaxy, in the direction of the steepest pressure decline.
The bubble enthalpy, the minimum energy required to produce a cavity, can be estimated by 
\begin{equation}\label{eq:7}
H=\frac{\gamma}{\gamma-1}PV
\end{equation}
where $P$ is the pressure, $V$ is the bubble volume, and $\gamma$ is the ratio of specific heats of the gas inside the cavity. 
Gas density at the radius of each cavity center is taken from the deprojected profile (Figure~\ref{fig:pro}), gas temperature is measured at the center of each cavity, and we apply $\gamma=4/3$ for relativistic gas. We use two ellipses (ellipsoids in 3D) to approximate the shapes of the cavities; their major axes are along the north-south directions (marked in Figure~\ref{fig:img}-bottom-right). 
We obtain $H=8\times10^{55}$\,ergs for the northern lobe and $H=1.2\times10^{56}$\,ergs for the southern lobe. 
We determine that the bubbles are in a subsonic expansion phase (\S3.4). 
Churazov et al.\ (2001) gives the terminal velocity of a buoyantly-rising bubble as
\begin{equation}\label{eq:8}
v=\sqrt{g\frac{2V}{SC}},
\end{equation} 
where $V$ and $S$ are the volume and cross section of the cavity respectively; $C\approx0.75$ is the drag coefficient.
The lobes would then rise at $490\rm\ km\ s^{-1}$, approximately 85\% of the sound speed, $c_s=\sqrt{\gamma kT/m_p\mu}$, with $\gamma=5/3$. 

{Both cavities appear more elongated along the jet-axis when compared to typical X-ray cavities. 
Many factors could affect the bubble shape (e.g., jet power, expansion phase, cluster mass). We observe elongated cavities in many different cases (e.g.,  M87-- Forman et al.\ 2007; Cygnus-A--Wilson et al.\ 2006; MS0735.6+7421--McNamara et al.\ 2005).  However, if the bubbles are too far from the plane of the sky (i.e., aligned along the line of the sight), they would appear rounder.} 
{We derive the expected X-ray surface brightness for each cavity 
by integrating the deprojected density profile along the line-of-sight for a range of different cavity positions on the line-of-sight.
We can examine whether a cavity is close to the
plane of the sky by comparing these expected values to the measured surface brightness.
As shown in Figure~\ref{fig:cavity}, the X-ray surface brightness of the northern lobe is consistent with what we expect for a cavity residing in the same plane of the sky as the nucleus while that of the southern lobe is higher than this expectation.
We attribute this discrepancy to the complicated filamentary structure surrounding the southern lobe.
It is therefore plausible for this pair of lobes to be in the plane of the sky that contains the nucleus.} 
Considering that they are 6--7\,kpc away from the nucleus, we estimate that they are about 14\,Myr old, corresponding to an outburst power of 2--3$\times10^{41}$\,erg\,s$^{-1}$ for each lobe. The properties of these lobes are summarized in Table~3.

We note the presence of a larger X-ray cavity at $\approx10$\,kpc in projection to the northwest of the nucleus, indicated in Figure~\ref{fig:N1399} and also visible in Figures~\ref{fig:img}-top-left. This surface brightness decrement was previously suggested by the ROSAT observations (Paolillo et al.\ 2002). It may be a radio cavity whose radio emission has faded and
dropped below the detection threshold as the bubble expanded and aged.
In this case, we estimate an enthalpy of $H=4\times10^{56}$\,ergs.  
Such ghost cavities are likely to be produced in previous AGN outbursts and they usually reside at larger radii than newly formed radio lobes. {Its surface brightness decrement translates to a distance of 9\,kpc from the plane of the sky (Figures~\ref{fig:cavity}) (13\,kpc to the nucleus). Adopting a terminal velocity of $v=\sqrt{g{2V}/{SC}}=366\rm\ km\ s^{-1}$, we infer that this ghost cavity is 34\,Myr old, corresponding to an outburst power of $3.6\times10^{41}$\,erg\,s$^{-1}$.} 
This cavity also resides in a different direction from the north-south axis implying that the spin axis of the blackhole may have rotated (e.g., RBS~797 - Cavagnolo et al.\ 2011) or the ghost bubble has been pushed around by the gas motion in the ambient medium (Abell~3581 - Canning et al\. 2013).  
Alternatively, this deficit in the X-ray surface brightness could be a consequence of being at the end of the spiral structure of the bright sloshing front. Gas sloshing in NGC~1399 is the subject of a companion paper (Su et al.\ submitted). 

\subsection{Spectroscopic maps}

We performed a two-dimensional spectroscopic analysis using Weighted Voronoi Tesselation (WVT) binning (Diehl \& Statler 2006)
based on the Voroni binning algorithm presented in Cappellari \& Copin (2003). We generated a WVT binned image containing 137 regions for NGC~1399 in the 0.5-2.0 keV band. Each bin has a S$/$N of 80 and contains approximately 3000 net counts.  
We use a single thermal {\tt vapec} component to model the hot gas emission in each region. 
The spectra were fit to the model {\tt phabs}$\times$({\tt vapec}+{\tt pow$_{1.6}$}). 
The resulting temperature map is shown in Figure~\ref{fig:img} (bottom-right). 
Cluster gas along the north-south directions is cooler than that in other directions out to 10\,kpc. The distribution of the cool gas follows the AGN jets and lobes, suggesting that low entropy gas from near the cluster center
has been carried outward by the AGN outburst.

\subsection{Bright rims and filaments}

Rims of enhanced X-ray surface brightness relative to the ambient cluster gas are visible adjacent to each cavity (rim1 and rim2 in Figure~\ref{fig:N1399}). The bright features could be shells of shocked gas during the supersonic expansion of the AGN bubbles (Forman et al.\ 2007). Alternatively, dense and cool gas from the cluster center can be lifted up by buoyant bubbles (Churazov et al.\ 2001; Gendron-Marsolais et al.\ submitted). As shown in the temperature map (Figure~\ref{fig:img}-bottom-right), the absence of heating in these regions strongly favors a subsonic phase of expansion. 
We approximate the shapes of the rims by an annular sector and a box region\footnote{For an annular sector, we calculate its volume as $V=\frac{4\pi}{3}({r_{\rm out}}^2-{r_{\rm in}}^2)^{1.5}\frac{\Delta\theta}{2\pi}$. For a box region, we assume it is a cylinder in 3D with an approximately radial symmetry axis and we calculate its volume as $V=l\times\pi(w/2)^2$.}, respectively, as marked in Figure~\ref{fig:N1399} and we extract spectra from these regions. 
{We fit the spectra of rims and filaments with the model {\tt phabs}$\times${\tt vapec}.  Their gas metallicities are fixed to the solar abundance. This single-temperature thermal model already gives very good fits ($\chi^2/d.o.f\sim1$). We do not have enough counts to perform a two-temperature fit and their irregular shapes are not suitable for deprojection analysis.  
To minimize the effect of the ambient hot cluster gas in projection, we adopt a local background extracted from a region in the relatively undisturbed ICM but at a similar radius as the rims to the cluster center.} 
As listed in Table~3, we measure a total gas mass of $\sim2\times10^{7}$M$_{\odot}$ in the rims.

In addition to the cool gas uplifted in front of the bubbles, 
we observe that cool gas has also been dragged out by the rising bubbles from the cluster center. As shown in the temperature map (Figure~\ref{fig:img}-bottom-right),
cool gas is present in the wake of the rising bubbles along the north-south directions. We observe a few bright filaments wrapping the bottom of the bubbles (f1, f2, f3 in Figure~\ref{fig:N1399}). We performed a similar spectral analysis for the filaments as for the rims. We measure a total gas mass of $\sim1.4\times10^{7}$M$_{\odot}$ in the filaments.
The properties of these bright rims and filaments are summarized in Table~3. 
Their presence implies that the minimum energy ejected by the jets should be the sum of the bubble enthalpy ($H$) and the work required to lift the gas ($W$).

 \begin{deluxetable*}{l|lll}

 \centering
\tablecaption{Properties of lobes, rims, and filaments}
\startdata
\hline
\\
{\bf Lobes}&north &south&ghost?\\
\hline
Distance traveled by lobe (kpc)&6.7 &7.1 & 13.0\\
Temperature at lobe center (keV) & 1.22$^{+0.05}_{-0.07}$ & 1.32$^{+0.02}_{-0.03}$ &$1.38^{+0.11}_{-0.03}$\\
Bubble enthalpy ($10^{55}$\,ergs) &$8$ &$12$&$40$\\
Rising speed (km/s) &504 & 481& 366 \\
Sound speed (km/s) & 568&583&597\\
Volume (kpc$^3$) &26.9 &39.4 &252.0 \\
Age (Myr)& 13.0& 14.4&34.0 \\
Power ($10^{41}$ erg/s)&1.9&2.5&3.6\\
Radio luminosity \footnotemark[1] ($10^{38}$ erg/s)&2.2&2.5&-\\
Minimum radio pressure\footnotemark[1] ($10^{-12}$ dynes/cm$^{2}$)& 2.8&2.5&-\\
Ambient X-ray pressure ($10^{-11}$ dynes/cm$^{2}$)& 2.5&2.5&1.3\\
\hline
\\
{\bf Rims}&north &south&\\
\hline
Current distance to cluster center (kpc)&8.5&10.4&\\
Initial distance to cluster center (kpc)&3.6&5.6&\\
Volume (kpc$^3$) &15.0 &52.7 &\\
Uplift mass ($10^6 \rm M_\odot$)&$6.8$ &$14.0$&\\
Uplift energy ($10^{54}$ ergs)&4.65&3.65&\\
Temperature (keV) & $1.23^{+0.03}_{-0.03}$& $1.33^{+0.02}_{-0.02}$&\\
$n_e$ ($10^{-3}$\,cm$^{-3}$) & $15.94^{+0.35}_{-0.45}$&$9.17^{+0.24}_{-0.19}$ &\\
Entropy (keV\,cm$^{2}$)&20 & 31&\\
\hline
\\
{\bf Filaments}&f1&f2&f3\\
\hline
Current distance to cluster center (kpc) &4.9&5.2&5.4\\
Initial distance to cluster center (kpc)&2.7& 2.8&2.5\\
Volume (kpc$^3$) & 6.9&6.5 &12.2\\
Uplift mass ($10^6 \rm M_\odot$)&$3.5$ &3.5 &$6.9$\\
Uplift energy ($10^{54}$ ergs)&$1.55$&$1.68$ &4.68\\
Temperature (keV) & $1.01^{+0.02}_{-0.02}$&$1.08^{+0.02}_{-0.02}$ &$1.02^{+0.01}_{-0.01}$\\
$n_e$ ($10^{-3}$\,cm$^{-3}$) & $17.94^{+0.57}_{-0.59}$&$19.03^{+0.36}_{-0.52}$ &$20.16^{+0.36}_{-0.33}$\\
Entropy (keV\,cm$^{2}$)&14.9 &15.4& 14.0\\
Cooling time (Gyr)&0.55 &0.60&0.51 
\enddata
\footnotesize{
\footnotetext[1]{taken from Killeen et al.\ (1988). The ratio ($k_p$) of pressure in other particles to that of relativistic particles is assumed to be unity.}}
\end{deluxetable*}

\section{\bf Discussion}

We performed a deep {\sl Chandra} analysis of NGC~1399, the BCG of the Fornax Cluster. We studied its hot gas properties within a radius of 25\,kpc, where {the hot X-ray gas is being influenced by AGN outbursts}. {Its X-ray hydrostatic mass profile approximates that of an isothermal potential with $M(r)\propto r$. The cluster cool core resides between a pair of relativistic plasma bubbles that are surrounded by rims and filaments of low entropy gas.} 
We discuss the cooling and heating processes at the center of this low-mass cool-core cluster. 

\begin{figure}
   \centering
       \includegraphics[width=0.5\textwidth]{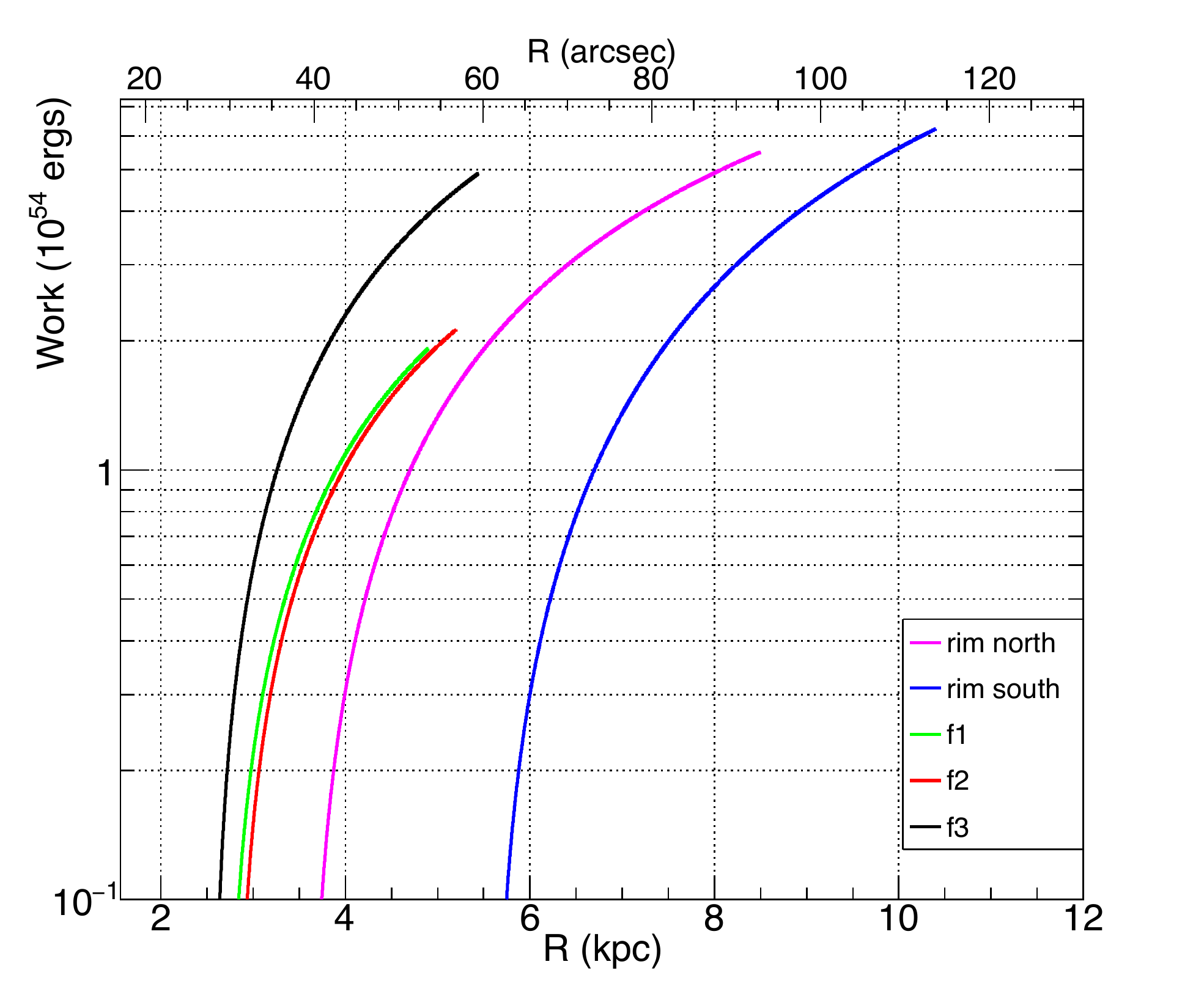} 
\figcaption{\label{fig:work} The minimum work required to lift the gas in the rims and filaments from their initial locations to their current positions.}
\end{figure}
{
\subsection{A quasi-isothermal potential}

The gravitational potentials of massive ellipticals are not far from being isothermal (Churazov et al.\ 2008, 2010).  
We derive the gravitational acceleration of NGC~1399 as a function of radius based on the hydrostatic equilibrium of hot gas (Figure~\ref{fig:pro}-right-bottom). Within a radius of 10\,kpc, the profile can be described by $g(r)={v_{\rm c}}^2/{r}$, as expected for an isothermal potential. It starts to deviate from this relation beyond $r=10$\,kpc (albeit with large uncertainties). This deviation may indicate a transition from the potential of the BCG to that of the Fornax Cluster. Hogan et al.\ (2017) also found that the NFW potential generally starts to be significant beyond about 10 kpc.

The circular velocity derived from X-ray data, $v_{\rm c}$, is $\lesssim10\%$ smaller than that from optical data, $v_*$. 
This discrepancy can be used to constrain the level of non-thermal pressure support. 
Following Churazov et al.\ (2010) and assuming that stellar dynamics (optical data) gives the unbiased mass estimate, 
we apply
\begin{equation}\label{eq:9}
{v_{*}}^2={v_{\rm c}}^2\left(1+\frac{P_{\rm n}}{P_{\rm t}}\right),
\end{equation}
where $P_{\rm n}$ and $P_{\rm t}$ are the non-thermal and thermal pressures, respectively. We obtain a $P_{\rm n}$ of $\lesssim10\%P_{\rm t}$ for the inner 10\,kpc; the non-thermal fraction places a constraint on the mechanism of AGN feedback in NGC 1399.
Assuming the power of AGN outburst balances radiative cooling, 
the respective energy densities of the thermal and non-thermal components can be related to the dissipation time of the mechanical energy and the cooling time scale (Churazov et al.\ 2008). 
We approximate the dissipation time by the age of the bubbles in NGC~1399 of 14\,Myr. The cooling time at the radii of the bubbles is $\gtrsim1$\,Gyr. Their ratio of $P_{\rm n}/P_{\rm t}\approx t_{\rm age}/t_{\rm cool}\approx10\%$ is in fully consistent with our estimate of $P_{\rm n}/P_{\rm t}$ using Equation~\ref{eq:9}. 
}

\subsection{Uplift of cluster gas by radio lobes}
We consider the scenario that the gas in the bright rims and filaments has been lifted by the buoyant bubbles inflated by the AGN outburst. The age of the bubbles is much shorter than the local cooling time. Therefore, the cool gas was unlikely to be heated or radiatively cooled appreciably while it was being lifted. Assuming the gas remains adiabatic and the mixing with the surrounding gas is not significant,  
we estimate the minimum work required to lift the gas.
Taking buoyancy into account, the net weight of an overdense gas blob is
\begin{equation}\label{eq:10}
F=(\rho_c-\rho)V_cg=M_c(1-\rho/\rho_c)g,
\end{equation}
where $g$ is the gravitational acceleration, $\rho$ is the ambient gas density,  $\rho_c$ and $V_c$ are the density and volume of the gas blob, respectively, and $M_c=\rho_cV_c$ is its mass.
We assume the ideal gas law and that the blob is at the same pressure as the ambient gas such that $\rho(r)/\rho_c=(K_c/K(r))^{1/\gamma}$, where $\gamma=5/3$, $K(r)$ is the entropy index for the ambient gas at a radius $r$, and $K_c$ is that of the blob which remains constant. The minimum work required to lift the gas blob is equal to the work that must be done against gravity to lift it from its initial, equilibrium location, $r_i$, to its current position, $r_f$, that is 
$$
W=\int^{r_f}_{r_i} F(r)dr
$$
\begin{equation}\label{eq:11}
=M_c\int^{r_f}_{r_i}[1-\{K_c/K(r)\}^{3/5}]g(r)dr.
\end{equation}
The uplifted gas should have originated from the radius where gas with about the same entropy is located now. The entropies of the rims and the filaments are 15--30\,keV\,cm$^{-2}$. We thus infer that the cool gas in the rims and filaments has traveled for 2--5\,kpc based on the entropy profile; none of the gas is directly from the very center of the cluster. We apply the best-fit entropy and gravity acceleration profiles within a radius of 10\,kpc (black solid lines in Figure~\ref{fig:pro}-right-middle and right-bottom) to Equation~\ref{eq:11}. We then obtain the minimum work required to lift the rims and filaments as shown in Figure~\ref{fig:work}, totaling $2\times10^{55}$\,ergs. Given a total bubble enthalpy of $H=2\times10^{56}$ ergs, we find that the uplift energy ($W$) is only 10\% of the total outburst energy ($H+W$). The power of the AGN outburst is then $\sim5\times10^{41}$\,erg\,s$^{-1}$. We caution the reader that the conventional approach to estimate the minimum work
\begin{equation}\label{eq:12}
W=\frac{M_c{c_s}^2}{\gamma}{\rm ln}\left(\frac{\rho_i}{\rho_f}\right), 
\end{equation}
e.g., employed in Reynolds et al.\ (2008), Gitti et al.\ (2011), and Kirkpatrick \& McNamara (2015), would overestimate the energy required to lift the gas (by 3--5$\times$) by discounting the buoyant force on the gas blob and assuming that the gas is lifted directly from the cluster center.

\subsection{Regulation of cooling}
While the cooling time drops below 1\,Gyr at the center of Fornax, as in
many massive cool-core clusters,
the star formation rate is low. We discuss the AGN mechanisms that may have prevented the hot gas from cooling catastrophically.  

{
The mechanical energy provided by AGN, mostly written in the enthalpy of bubbles, is expected to offset radiative gas cooling (e.g., B{\^i}rzan et al.\ 2004; Rafferty et al.\ 2006). 
We infer a total cavity power of $P_{\rm cav}=4.4\times10^{41}$\,erg\,s$^{-1}$ for NGC~1399 (Table~3). This value is almost the same as the bolometric X-ray luminosity within the cool core (Table~2).
B{\^i}rzan et al.\ (2008) calibrated a scaling relation between
the cavity power, $P_{\rm cav}$, and the radio
power at 1.4GHz, $P_{\rm 1400}$, for galaxy clusters:
\begin{equation}\label{eq:13}
{\rm log}P_{\rm cav}=0.35(\pm0.07) {\rm log}P_{\rm 1400}+1.85(\pm0.10).
\end{equation}
O'Sullivan et al.\ (2011) determined a similar relation
for a sample of galaxy groups: 
\begin{equation}\label{eq:14}
{\rm log}P_{\rm cav}=0.63(\pm0.10) {\rm log}P_{\rm 1400}+1.76(\pm0.15).
\end{equation}
$P_{\rm cav}$ and $P_{\rm 1400}$ in the above two equations are in units of $10^{42}$\,erg\,s$^{-1}$ and $10^{24}$\,W\,Hz$^{-1}$, respectively.   
For a $P_{\rm 1400}$ of $2.1\times10^{22}$\,W\,Hz$^{-1}$ in NGC~1399 (Shurkin et al.\ 2008), 
we expect $P_{\rm cav}$ to be $1.8\times10^{43}$\,erg\,s$^{-1}$ (Equation~\ref{eq:13}) and $5\times10^{42}$\,erg\,s$^{-1}$ (Equation~\ref{eq:14}), respectively. That we measure a $P_{\rm cav}$ of $<1\times10^{42}$\,erg\,s$^{-1}$
implies that the coupling between the mechanical and synchrotron luminosities of the radio source in NGC~1399 is weak when compared to other galactic systems, although a disagreement by a factor of 10 is typical among radio faint sources.}

In addition (or as an alternative) to the direct re-heating paradigm, we consider the scenario that buoyant bubble may prevent gas from cooling by lifting cool materials from the cluster center. 
We calculate the nominal cooling rate of the gas through
\begin{equation}\label{eq:15}
\dot{M}_{\rm cool}=\frac{2}{5}\frac{\mu m_pL_X}{kT_X},
\end{equation}
where $L_{\rm X}$ is the bolometric X-ray luminosity (listed in Table~2). 
Integrating within a radius of 25\,kpc, we obtain a $\dot{M}_{\rm cool}$ of 1.4\,M$_{\odot}$\,yr$^{-1}$ in spite of an observed star formation rate of $\ll0.1$\,M$_{\odot}$\,yr$^{-1}$ (Vaddi et al.\ 2016). 
We observe a total of $3.4\times10^{7}$\,M$_{\odot}$ cool gas in the rims and filaments, corresponding to an uplift rate of $\dot{M}_{\rm uplift}=2.5$\,M$_{\odot}$\,yr$^{-1}$ (Table~3). Thus 
all the cool gas that is expected to fuel cooling flows and star formations can be carried away from the cluster center by AGN uplift. 
This is in contrast with more massive clusters where 
the uplift rate can account for only 10--20\% of the cooling rate and it is thus insufficient to prevent gas from cooling (Kirkpatrick \& McNamara 2015; but see Gitti et al.\ 2011). {Kirkpatrick \& McNamara (2015) found that the uplift rate scales with the cavity power  
\begin{equation}\label{eq:16}
\dot{M}_{\rm uplift}=(22\pm16)\times {P_{\rm cav}}^{1.4\pm0.4} ~~~\rm M_{\odot}\,yr^{-1},
\end{equation}
where $P_{\rm cav}$ is in units $10^{44}$\,erg\,s$^{-1}$. Based on this crude relationship, we expect an uplift rate of only $\dot{M}_{\rm uplift}=0.011~\rm M_{\odot} yr^{-1}$ for NGC~1399, while we observe $\dot{M}_{\rm uplift}=2.5$\,M$_{\odot}$\,yr$^{-1}$, implying that AGN bubbles in NGC~1399 are efficient in lifting cool materials.
These differences may reflect the ICM in low-mass clusters being more sensitive to AGN feedback thanks to their shallower gravitational potentials (Giodini et al.\ 2010).}

We argue that uplift can only help alleviate the radiative gas cooling as opposed to quench the cooling flow by itself. 
In NGC~1399, the observed rims and filaments are lifted from radii of 2--5\,kpc, instead of from the innermost region. {Uplifted cool gas would eventually fall back (albeit to a radius larger than its original altitude) and may be unable to eventually leave the cool core}. 
McNamara \& Nulsen (2012) has also demonstrated that adiabatic uplift can delay gas cooling by no more than a factor of 3; cooling cannot be prevented by uplift alone.

\subsection{The single phase ICM}

Molecular cold gas has been detected in a growing number of cool-core clusters (David et al.\ 2014; Russell et al,\ 2014, 2016; McNamara et al.\ 2014; Edge \& Frayer 2003). 
This is generally believed to be the
product of thermally unstable cooling from the hot ICM (McCourt et
al.\ 2012; Voit \& Donahue 2015; McNamara et al.\ 2016), although other
sources, such as merger debris from gas-rich galaxies cannot be ruled
out. Recent simulations suggest that cold gas is expected to form once  
the ratio of the cooling time scale $t_{\rm cool}$ and the dynamical time scale $t_{\rm ff}$ drops below {a certain (yet indeterminate) value}.
NGC~1399 lacks such molecular cold gas (Werner et al.\ 2014). 
We derive the $t_{\rm cool}/t_{\rm ff}$ ratio for its central region as shown in Figure~\ref{fig:pro}, using the X-ray hydrostatic mass profile.  
It stays above 30 over the entire cluster center. We inspect the local $t_{\rm cool}/t_{\rm ff}$ ratio for the bright filaments and still obtain a large value. 
We also use the total mass profile derived from the globular cluster kinematics (Samurovic \& Danziger 2006) and obtain $t_{\rm cool}/t_{\rm ff}=70$ and $t_{\rm cool}/t_{\rm ff}=110$ within radii of $2^{\prime}$ and $4^{\prime}$, respectively.
In contrast, massive galaxy clusters that do contain cold gas seem to have a minimum $t_{\rm cool}/t_{\rm ff}$ ratio below 20 (Hogan et al.\ 2017). Our results are consistent with thermal
instability being controlled by the ratio $t_{\rm cool}/t_{\rm ff}$ in low mass
clusters as well as the more massive clusters. {It is noteworthy that (non-linear) density perturbations (e.g., triggered by outflows or cavities) can in principle give rise to cold gas even at $t_{\rm cool}/t_{\rm ff}=60$ as found in simulations (Valentini \& Brighenti 2015).}

\subsection{Supernova driven outflow}

\begin{figure}
   \centering
    \includegraphics[width=0.5\textwidth]{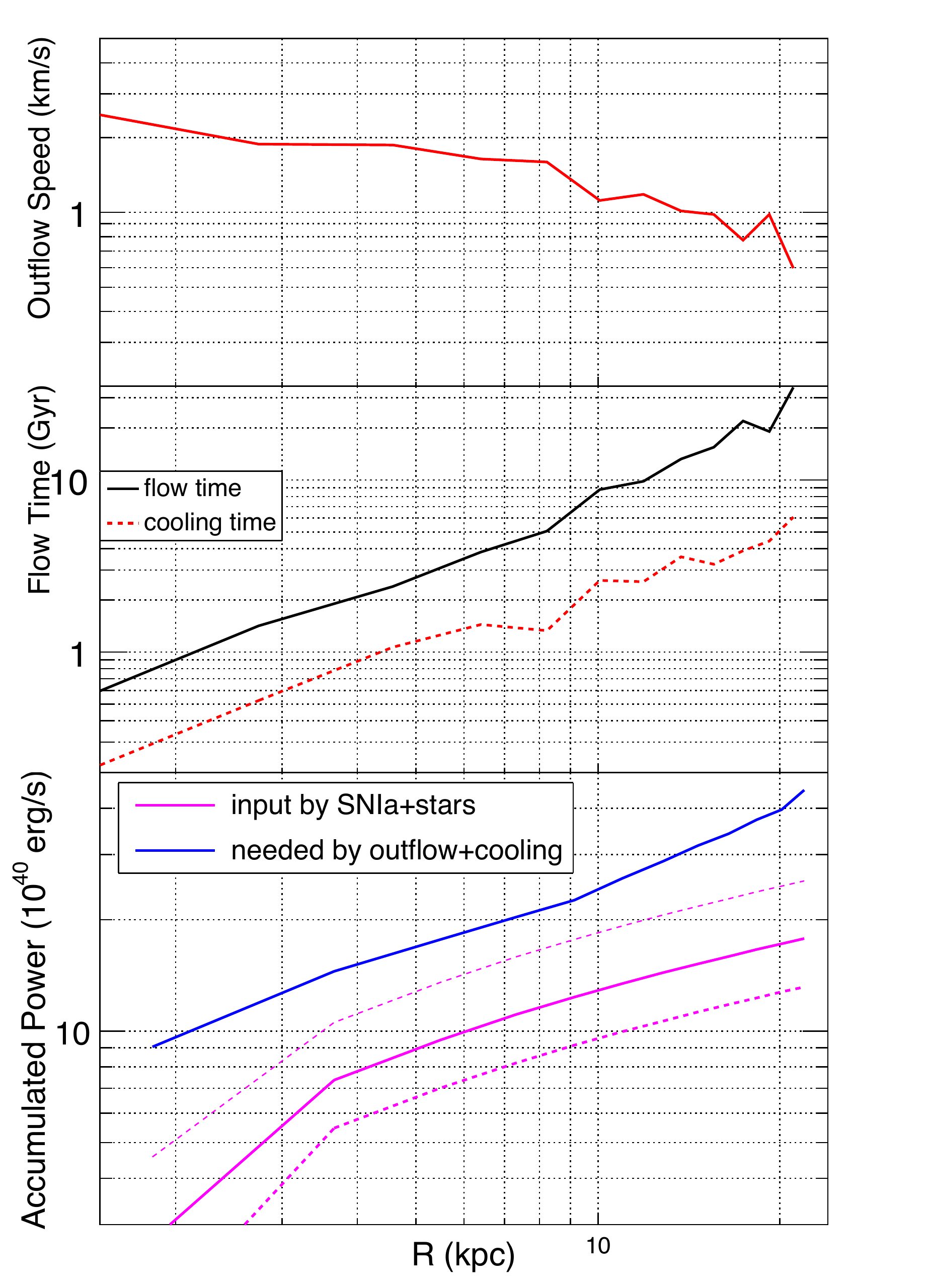}
\figcaption{\label{fig:wind} {\it top}: Outflow speed at each radius. {\it middle}: The time for a slow outflow to reach radius $r$ (black), compared to the local cooling time (red). {\it bottom}: The energy injection rate by SNIa and old stars within $r$ (magenta); dashed lines characterize the uncertainty in the delay time distribution for SNIa. Blue solid line indicates energy loss rate in the hot gas within $r$.}
\end{figure}

The entropy profile of NGC~1399 deviates from that of massive clusters and the prediction for  purely-gravitational structure formation (Figure~\ref{fig:pro}-middle-right). 
Instead, it follows the entropy profile of single-phase elliptical galaxies ($K=5\,(r/{\rm kpc})$\,keV\,cm$^2$, Voit et al.\ 2015). 
The cooling and heating processes of multi-phase systems are regulated by thermal instability and responsive AGN feedback (Revaz et al.\ 2008; Brighenti et al.\ 2015). In contrast, these of single-phase systems may be governed by Type I Supernova (SNIa) driving outflow.
We examine the scenario in which hot gas in NGC~1399 is shed by stars and energized by SNIa.

We assume a time scale of $\tau=10^{12}$ years for the stellar mass loss. The mass injection rate per unit volume is then $\alpha=\rho_{\ast}/\tau$. To determine the stellar distribution, we derive the $K$-band luminosity of NGC~1399 using the {\sl Two Micron All Sky Survey} ({\sl 2MASS}) (Skrutskie et al.\ 2006). Procedures of the {\sl 2MASS} data reductions are presented in Su et al.\ (2013, 2015). NGC~1399 is an E0-1 type early-type galaxy for which we assume spherical symmetry. A luminosity density $j^{\rm lum}(r)$ can be derived from the surface brightness profile $\Sigma(R)$ using the standard Abel equation 
\begin{equation}\label{eq:17}
j^{\rm lum}(r)=-\frac{1}{\pi}\int^{\infty}_{r}\frac{d\Sigma(R)}{dR}\frac{dR}{\sqrt{R^2-r^2}}.
\end{equation}
The spherically integrated $K$-band luminosity of NGC~1399 is $2.1\times10^{11}L_{\rm K,\odot}$ within a radius of 20\,kpc. We adopt a $K$-band mass-to-light ratio of $M_{*}/L_{\rm K}=1.3$\,M$_{\odot}/L_{\rm K, \odot}$ for 10\,Gyr old stars (Silva \& Bothun 1998). 
From $j^{\rm lum}(r)$ we calculate the total stellar mass within $r$ of $M_{*}(r)=\int_0^r 1.3 j^{\rm lum}(r') 4 \pi r'^2 dr'{\rm M_{\odot}}/L_{\rm K,\odot}$.

For a steady slow supernova heated wind, the outward mass flow rate though a shell of radius $r$ is
\begin{equation}\label{eq:18}
\dot M(r)=4\pi\rho v r^2=\int_{0}^{r}4\pi\alpha(r^{\prime}){r^{\prime}}^2 dr^{\prime}=M_{*}(r)/\tau.
\end{equation}
The flow velocity at each radius is then 
\begin{equation}\label{eq:19}
v = \dot{M}/(4\pi \rho r^2)
\end{equation}
Its value lies below 5\,km\,s$^{-1}$ (slow outflow) over the entire cluster center as shown in Figure~\ref{fig:wind}-top. 
Consequently, we estimate the time for the flow to reach each radius as
\begin{equation}\label{eq:20}
t_{\rm flow}(r)= \frac{r}{v} = 4\pi \rho r^3 / \dot{M}.
\end{equation}
As shown in Figure~\ref{fig:wind}-middle, the flow time stays above the cooling time. 
It is therefore far-fetched for the supernova driven outflow to compensate for the radiative cooling.

The delay time distribution for SNIa is 
\begin{equation}\label{eq:21}
\Psi(t)=\Psi_{\rm 1\,Gyr}(t/{\rm 1\,Gyr})^{-1},
\end{equation}
where $\Psi_{\rm 1\,Gyr 
}\approx1.7^{+0.5}_{-0.8}\times10^{-13}\,\rm SNIa\,{M_{\odot}}^{-1}\,yr^{-1}$ (Graur et al.\ 2014).
We obtain $\Psi=1.7\times10^{-14}\,\rm SNIa\,{M_{\odot}}^{-1}\,yr^{-1}$ for $10$\,Gyr old stellar population. 
Adopting an energy deposition per SNIa event of $E_{\rm SN}=10^{51}$\,erg, the supernova energy input per unit stellar mass loss is $\epsilon_{\rm SN}=\Psi\tau E_{\rm SN}$. 
Another heating source is the kinetic energy of the winds
from old stars, which is dominated by the random motions of the stars;
the energy input per unit stellar mass loss is $\epsilon_{\ast}=\frac{3}{2}{{\sigma_{*}}^2}$; for NGC~1399, we use $\sigma_{*}=300$\,km\,s$^{-1}$ (Saglia et al.\ 2000).
The integrated energy equation for the steady
flow relates the total energy outflow through a shell of radius $r$,
\begin{equation}\label{eq:22}
 \dot{M}[H(r) + \frac{1}{2} v^2 + \Phi(r)],
\end{equation}
to the net rate of energy input within the shell. 
The rate of energy injection within the shell due to SNIa and old
stars is
\begin{small}
\begin{equation}\label{eq:23}
\int_{0}^{r}4\pi\alpha(r^{\prime})[\epsilon_{\rm SN}+\epsilon_{*}+\Phi(r^{\prime})]{r^{\prime}}^2dr^{\prime},
\end{equation}
\end{small}
while the rate of energy loss due to radiation is
\begin{small}
 \begin{equation}\label{eq:24}
 \int^{r}_{0}4\pi {r^{\prime}}^2n_e(r^{\prime})\Lambda[T(r^{\prime}),Z(r^{\prime})]dr^{\prime}. 
\end{equation}
\end{small}
The term involving $\Phi$ in Equation~\ref{eq:23}, can be
integrated by parts to give
\begin{equation}\label{eq:25}
\int_0^r 4 \pi \alpha(r') \Phi(r') r'^2\,dr'
= \dot{M}(r) \Phi(r) - \int_0^r \dot{M}(r') g(r') \, dr'.
\end{equation}
The first term on the right here can be canceled with the
term involving $\Phi$ in Equation~\ref{eq:22}. We also
neglect the term $\frac{1}{2}v^2$ for the slow outflow.
Moving the
radiative loss to the other side of the equation, we find that the net
energy input rate related to SNIa and stars,
\begin{small}\label{eq:26}
$$P_{\rm input} (r) =\int_{0}^{r}4\pi\alpha(r^{\prime})[\epsilon_{\rm SN}+\epsilon_{*}]{r^{\prime}}^2dr^{\prime}$$
\begin{equation}
-\int_{0}^{r}\dot{M}g(r^{\prime})dr
\end{equation}
\end{small}
must balance the enthalpy outflow and radiative loss
\begin{small}
$$
P_{\rm needed}(r) = \dot{M}(r) H(r) 
$$
\begin{equation}\label{eq:27}
+\int_0^r 4 \pi r'^2 n_e(r')n_H(r') \Lambda[T(r'), Z(r')]\, dr'.
\end{equation}
\end{small}
As shown in Figure~\ref{fig:wind}-bottom, $P_{\rm needed}$ exceeds $P_{\rm input}$ over the entire cluster center even when the uncertainty in the SNIa delay time distribution is taken into account, as found in simulations (see Mathews \& Brighenti 2003 for a review).

In addition, this scenario cannot be easily reconciled with the metal content of hot gas. Since each SNIa event produces $M_{\rm Fe}=0.7$\,M$_{\odot}$ iron yield, we expect an iron mass fraction of $f_{\rm Fe}=\Psi\tau M_{\rm Fe}\simeq0.012$. This would correspond to a Fe abundance of $\sim5$\,Z$_{\odot}$. In contrast, the observed Fe abundance  in NGC~1399 is $\lesssim1$\,Z$_{\odot}$.
If the iron yield were not completely incorporated into the gas being shed by the stars, then we expect a large reservoir of dust must have absorbed the metals. However, few dust features are detected in NGC~1399 (Prandoni et al.\ 2010; Davies et al.\ 2013); metals are unlikely to be hidden in the dust. Heating supplied by SNIa and old stars is therefore insufficient to drive slow outflow, compensate for cooling, and maintain a single-phase ICM.

\section {\bf Conclusions}

The Fornax Cluster is a nearby low-mass cool-core cluster. It harbors a pair of symmetric radio lobes coincident with two X-ray cavities. 
We have analyzed a total of 250\,ksec {\sl Chandra} observations centered on its BCG NGC~1399 and present the thermal properties of the hot gas within a radius of 25\,kpc (0.05\,$R_{\rm 500}$).  
We find that: 
\medskip

 $\bullet$ The X-ray cavities are 6--7\,kpc away from the nucleus. We estimate that the AGN bubbles are about 14\,Myr old corresponding to a total AGN outburst power of $5\times10^{41}$\,erg\,s$^{-1}$. 
We note a potential ghost bubble of $\gtrsim$ 5\,kpc in diameter to the northwest with an enthalpy of $4\times10^{56}$\,ergs and an age of 34 Myr. 

$\bullet$ The X-ray cavities are surrounded by rims and filaments of enhanced surface brightness that are cooler than the ambient ICM. We infer that they consist of low-entropy gas lifted from smaller radii (instead of directly from the nucleus) by the buoyant AGN bubbles. 
The minimum energy required to lift the cool gas against gravity is a small fraction ($10\%$) of the energy released by the AGN outburst.  

 $\bullet$ Cool gas uplifted by AGN bubbles at a rate of 2.5\,M$_{\odot}$\,yr$^{-1}$ can account for all of the gas that is expected to cool catastrophically, while uplift in massive clusters can only remove 10--20\% of the gas that cools in their cores.

 $\bullet$ The cooling time of this cluster is more than 30$\times$ longer than the dynamic time over the entire cluster center. This is consistent with the lack of cold molecular gas in this system.

$\bullet$ Outflow driven by SNIa events is insufficient to produce and maintain the thermal distribution in NGC~1399. It is also in tension with the iron content observed in the hot gas.

\section{\bf Acknowledgments}
This work was supported by Chandra Awards GO1-12160X and GO2-13125X issued by the Chandra X- ray Observatory Center which is operated by the Smithsonian Astrophysical Observatory under NASA contract NAS8-03060.

\end{document}